\documentclass[a4paper,11pt]{article}
\pdfoutput=1 % if your are submitting a pdflatex (i.e. if you have
\usepackage{jheppub} % for details on the use of the package, please
\usepackage[T1]{fontenc} % if needed
\usepackage{soul}
\usepackage{verbatim} 
\usepackage{tikz}
\usetikzlibrary{decorations.markings,calc,shapes,decorations.pathmorphing}
\usetikzlibrary{patterns}
\usetikzlibrary{positioning}

\title{\boldmath Double brane holographic model dual to 2d ICFTs}

\author[a]{Saba Asif Baig}
\author[a]{, Andreas Karch}

\affiliation[a]{University of Texas at Austin, Physics Department,\\ Austin TX 78712, USA}

\emailAdd{karcha@utexas.edu}
\emailAdd{sbaig@utexas.edu}

\abstract{A minimal single brane holographic model can be used as a dual to 2d conformal interfaces (ICFTs) to calculate the transmission coefficient $\mathcal{T}$ of energy transported across the defect as well as boundary entropy $\log g$, the additional entanglement entropy for some sub-region that encloses the defect. Both $\mathcal{T}$ and $\log g$ are uniquely determined by the tension characterizing the brane. In contrast, in field theory defects typically the transmission coefficient can be dialed from 0 to 1 independently for each allowed value of $\log g$. To address this discrepancy, we look at a double brane (3-region bulk) holographic model. Merger of two single brane interfaces creates genuinely new interfaces which indeed allow a range of accessible transmission coefficients for a fixed value of $\log g$. In particular, the $\mathcal{T}=0$ limit of two completely decoupled BCFTs can be achieved.}

\begin{document} 
\maketitle
\flushbottom

\section{Introduction}

Conformal interfaces refer to the local gluing of two scale-invariant conformal field theories, CFT$_{L}$ and CFT$_{R}$. Such interfaces are well studied in condensed matter physics, in describing junctions~\cite{a}, defects in Ising Models~\cite{b}, etc, as well as in the string theory literature \cite{p}. This work considers 1+1 dimensional CFTs where the interface is a point-like defect, and bottom-up AdS/CFT correspondence gives the dual holographic model as an AdS$_3$ geometry with a codimension-one brane representing the defect, anchored at the boundary~\cite{d,e,q}. CFT$_{L}$ and CFT$_{R}$ have central charges $c_{L}$ and $c_{R}$ respectively, and their corresponding holographic duals AdS$_{3,L}$ and AdS$_{3,R}$ have radii $l_{L}$ and $l_{R}$. The Brown-Henneaux formula $c_{L,R} = 3l_{L,R}/2G$ ~\cite{l} relates the two where $G$ is the 3 dimensional Newton's constant. 

\smallskip
Recent studies use this holographic model to evaluate two properties for 2d ICFTs: a transmission coefficient of energy transported across the defect~\cite{f}, and entanglement entropy for some sub-region of length a, that encloses the defect~\cite{n,g}. Our work probes this defect further by looking at the behaviour of these properties under the fusion of two point-like defects, in the limit that they lie on top of one another. This gives a wedge-shaped bulk with AdS$_3$ radius $l_{C}$ for the 'centre' region between the two defects. Our double brane holographic model gives closed form expressions for $\mathcal{T}$ and $\log g$ that clearly show the merger of two single brane defects can not be described by another single brane defect. There is no closed algebra to describe the fusing behaviour for these general single brane 2d ICFTs, unlike what is present in critical defects in 2d Ising Models~\cite{h,i} and other well understood CFTs~\cite{j}. The allowed interfaces in an ICFT must form a closed algebra under fusion. Therefore, multi-brane defects can either be considered as genuinely new objects in the system or as a non-trivial quantum superposition of existing defects. We'll argue that the former is this case and so fusion only has a chance of closing in this extended space of allowed interfaces. Presumably, the underlying reason for this failure of fusion to close on the single brane defects is the large $N$ limit that is required in any holographic theory. One would expect that quantum effects allow a multi-brane interface with $N$ branes, where $N$ is a large integer related to the central charge, to become equivalent to a single brane interface, very much like in the Ising model where a fusion of two branes with $g=\sqrt{2}$ does not yield a new brane with $g=2$ but instead a superposition of two separate $g=1$ theories.~\cite{h,i}. Such a purely quantum effect clearly will not be visible in our large $N$ classical limit holographic setting. The fact that multi-brane interfaces can be used to enlarge the space of allowed holographic interfaces has also been recently used in \cite{Ooguri:2020sua} in the context of the co-bordism conjecture.\cite{McNamara:2019rup}. Unlike in our case with a small intermediate region, they consider triple CFT interfaces.

\smallskip
We show that multi-brane interfaces permit the bottom-up holographic defect to sustain more tension and therefore allow increasingly more reflection while maintaining AdS$_3$ geometry. With 2 branes, the transmission coefficient can be dialed to 0 by setting $l_{C}$ to 0, regardless of the values of $l_{L,R}$. Increasing the number of defects fused together also enables imposing reflecting boundary conditions ($\mathcal{T} \rightarrow 0$) for the same AdS radius describing all regions. 

\smallskip
The paper is organized as follows. In section 2, we describe the scattering states in the double brane holographic model, compute the transmission coefficient and comment on its useful limits. In section 3, we use the constraints from the double brane model to evaluate the additional entanglement entropy log $g$ and present the results for a specific $\mathbb{Z}_2$ parity symmetry. In the last section, we conclude and discuss possible future work. Appendix A contains the details of the linear order differential system solved to obtain the transmission coefficient result and in appendix B, we present expressions for $\log g$ in the general double brane model. 

\section{Transmission coefficient for the double brane model}

\subsection{Review: Single brane toy model} 

In order to use established properties of B(oundary)CFTs, the ICFT can be mapped to a BCFT of the product theory CFT$_L$ $\otimes$ $\overline{\text{CFT}_R}$ by folding spacetime along the interface. The BCFT conserves energy momentum tensors T$_L$ and $\overline{\text{T}_R}$ separately in the bulk, and their sum on the boundary. It also conserves a relative spin-2 current T$_{rel} = c_R\text{T}_L - c_L\overline{\text{T}_R}$ that measures the relative exchange of energy across the interface between the two CFTs. In 2d, this exchange of energy can be expressed using a single real transmission or reflection coefficient~\cite{m}, which is computed using a toy holographic model.

\smallskip
The toy model is given by two AdS$_{3}$ slices separated by a string of tension $\sigma$~\cite{o,d,q,p}. The AdS$_{3}$ slices have radius $l_{L}$ and $l_{R}$. The string's worldsheet geometry is AdS$_{2}$, corresponding to the ground state of the ICFT for the following tension range~\cite{d}: 
\begin{equation}\label{oldtensionbound}
    \left| \frac{1}{l_{L}} - \frac{1}{l_{R}} \right| \leq 8\pi G \sigma \leq \frac{1}{l_{L}} + \frac{1}{l_{R}}
\end{equation}
The lower tension bound corresponds to the Coleman De Lucia bound~\cite{r}. Below this, the space with the higher AdS$_{3}$ radius ('false vacuum') is unstable to the nucleation of bubbles (which permit 'decays' between vacua). However, since for decays in AdS tunnelling is not always allowed (the false vacuum can be stable) or when allowed, the endpoint cannot be the true vacuum, this can't always be understood in terms of vacuum decay. Instead it is more useful to interpret it as holographic RG flows between conformal fixed points~\cite{t}. The upper tension bound corresponds to the Randall-Sundrum fine tuned tension beyond which the string worldsheet geometry becomes de-Sitter~\cite{s}.

\subsection{Fusion of interfaces: 3-region holographic model}
\begin{figure}[ht!]
\centering
\begin{tikzpicture}[scale=0.50]

% Horizontal line at top of 3 non-zero regions
\draw[<-,blue,thick] (-7.5,3) to (-4.5,3);
\draw[-,blue,thick] (-1.5,3) to (1.5,3);
\draw[->,blue,thick] (4.5,3) to (7.5,3);

% Vertical lines representing branes
\draw[-,blue,thick] (-4.5,3) to (-3.5,0);
\draw[-,blue,thick] (-2.5,0) to (-1.5,3);
\draw[-,blue,thick] (1.5,3) to (2.5,0);
\draw[-,blue,thick] (3.5,0) to (4.5,3);

% Co-ordinate lines (left, then right)
\draw[->,red!30!blue!90!, very thick] (-4.5,3) -- (-4.5,1);
\node[red!30!blue!90!] at (-5,1.5) {$y_{L}$};
\draw[->,red!30!blue!90!, very thick] (-4.5,3) -- (-3,3);
\node[red!30!blue!90!] at (-3.5,3.5) {$u_{L}$};

\draw[->,red!30!blue!90!, very thick] (4.5,3) -- (4.5,1);
\node[red!30!blue!90!] at (5,1.5) {$y_{R}$};
\draw[->,red!30!blue!90!, very thick] (4.5,3) -- (6,3);
\node[red!30!blue!90!] at (5.5,3.5) {$u_{R}$};

% marking central region
\draw[<->,thin] (-1.5,0.5) to (1.5,0.5);
\node at (0,1) {$\delta$};

% Central region goes to single point at the boundary
\draw[->,thin] (0,-0.5) to (0,-2);
\node at (1.5,-1.5) {$\delta=0$};

% 3region model (left top, then brane, then angle)
\node[blue!70!] at (-9,-11) {$M_{L}$};
\node[red!70!] at (-9,-12) {$l_{L}$};

\draw[-,blue,very thick] (-10,-4) to (-6,-4) to (-7,-8) to (-11,-8) ;
\draw[->,red!50!blue!70!, thick, style={decorate, decoration={snake, post=lineto ,post length=0.1cm}}] (-10,-7.5) -- (-8,-6.5);
\node[red!50!blue!70!] at (-10.5,-7) {$I$};
\draw[->,red!50!blue!70!, thick, style={decorate, decoration={snake, post=lineto ,post length=0.1cm}}] (-8,-5.5) -- (-10,-4.5);
\node[red!50!blue!70!] at (-7.5,-5) {$R$};

\draw[-,blue, very thick, fill=red!10!, style={decoration=snake}] (-4,-9) -- (-6,-4) -- (-7,-8) -- (-5,-13) decorate{-- (-4,-9)};
\node[red!70!] at (-4,-13) {$\sigma_{1}$};

\draw[dashed, blue, thin] (-7,-8) to (-7,-11);
\draw[->, red, thick] (-7,-9.5) arc(-90:-68.2:1.5cm);
\node[red!70!] at (-6.5,-10) {$\theta_{L}$};

% 3region model (right top, then brane, then angle)
\node[blue!70!] at (9,-11) {$M_{R}$};
\node[red!70!] at (9,-12) {$l_{R}$};

\draw[-,blue,very thick] (10,-4) to (6,-4) to (7,-8) to (11,-8);
\draw[->,red!50!blue!70!, thick, style={decorate, decoration={snake, post=lineto ,post length=0.1cm}}] (7,-6) -- (9,-5);
\node[red!50!blue!70!] at (7.5,-5) {$T$};

\draw[-,blue, very thick, fill=green!10, style={decoration=snake}] (4,-9) -- (6,-4) -- (7,-8) -- (5,-13) decorate{-- (4,-9)};
\node[red!70!] at (4,-13) {$\sigma_{2}$};

\draw[dashed, blue, thin] (7,-8) to (7,-11);
\draw[->, red, thick] (7,-9.5) arc(-90:-112.8:1.5cm);
\node[red!70!] at (6.6,-10) {$\theta_{R}$};

% 3region model (centre, then left brane, then right brane, then angles )
\node[blue!70!] at (0,-12) {$M_{C}$};
\node[red!70!] at (0,-13) {$l_{C}$};

\draw[-,blue,very thick, fill=red!10!, style={decoration=snake}] (0,-4) -- (-3,-9) decorate{-- (-3,-13)} -- (0,-8) -- (0,-4);

\draw[-,blue,very thick, fill=green!10!, style={decoration=snake}] (0,-4) -- (3,-9) decorate{-- (3,-13)} -- (0,-8) -- (0,-4);

\draw[dashed, blue, thin] (0,-8) to (0,-11);
\draw[->, red, thick] (0,-9.5) arc(-90:-121:1.5cm);
\node[red!70!] at (-0.5,-10) {$\theta_{CL}$};
\draw[->, red, thick] (0,-10) arc(-90:-59:2cm);
\node[red!70!] at (0.8,-10.5) {$\theta_{CR}$};

\end{tikzpicture}
\caption{\label{fig1}The 3 region holographic model representing the fusion of two defects. I is the gravitational wave scattered from the left, and R and T are the reflected and transmitted waves on the left and right of the interface respectively.}
\end{figure}
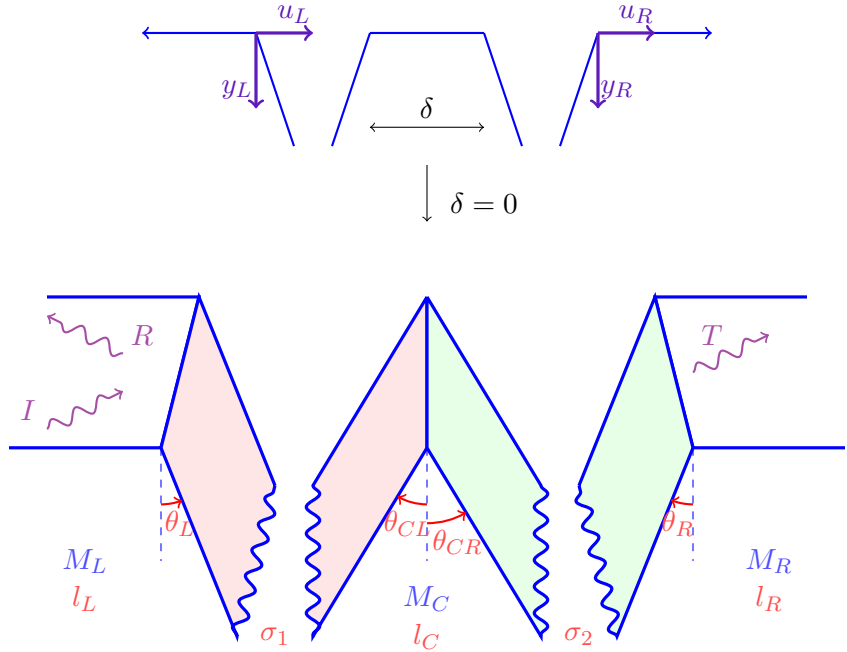

Consider two distinct interfaces (CFTs prescribed for three regions) and then shrink the intermediate region's size to 0 so the interfaces lie on top of one another. In this limit, the 1+1 dimensional ICFT can be represented by this new 3-region 'double brane' holographic model, as shown in Figure \ref{fig1}. The main idea is to check if this double brane model belongs in the same class as the single brane models. If fused interfaces belong to a different class, then this technique may help complete the classification of holographic defects and check if they compose under an algebra.

\smallskip
In a single defect holographic setting, the tension $\sigma$ uniquely determines $\mathcal{T}$ and $g$. In contrast, as emphasized in the introduction, in field theories all values of the transmission coefficient $\mathcal{T}$ are realizable for a given $g$~\cite{h}. $\log g$ is the additional entanglement entropy due to the presence of a defect within the given sub-system (of length a)~\cite{v,u,n}. For a symmetric interval, the entanglement entropy takes a form~\cite{n,q} that agrees with the standard BCFT result~\cite{u} for a BCFT with central charge $c_L + c_R$: 
\begin{equation}
    EE = \frac{c_L + c_R}{6} \log(\frac{a}{\epsilon}) + \log g
\end{equation}

\smallskip
We find that indeed fused interfaces are genuinely different from single brane defects, with an infinite number of classes possible, where the $g$ values increase unbounded and the transmission coefficient is given by the sum of the allowed tensions. This classification of defects is in contrast with the Ising model where $g$ values can be used to classify between Ising models with Neumann and Dirichlet boundary conditions, and fusion of defects follows the Verlinde algebra.

\subsection{Holographic scattering states}

We use a 3-region holographic model for the ICFT, consisting of three locally AdS$_{3}$ manifolds $M_{L}$, $M_{C}$ and $M_{R}$ with radii $l_L, l_C$ and $l_R$ respectively. The asymptotic boundaries of $M_{L}$ and $M_{R}$ are the left and right half-planes of the CFT glued along the interface. The asymptotic boundary of $M_{C}$ is a single line that gives the worldline of the same interface. The interface extends in the bulk as two pairs of branes and a central region $M_{C}$: the branes from $M_{L}$ and the left side of $M_{C}$ are identified with each other as the worldsheet of one tensile string and the branes from $M_{R}$ and the right side of $M_{C}$ are identified with one another and as the worldsheet of a second tensile string. This identification implies that the $M_{C}$ wedge always has a `non-negative length.' While solving (\ref{34}) which arises from these matching conditions, we obtain conditional solutions that give valid tension bounds. For an $M_{C}$ wedge found entirely on either side of the interface line, a 'non-negative length' requirement gives a lower bound on $\sigma_2$ which depends on $\sigma_1$ (or vice versa) and, while it can be written in closed form, it does not appear illuminating. Therefore, going forward, we will only present scenarios where the wedge extends on both sides of the interface line and we can write down independent bounds for $\sigma_1$ and $\sigma_2$.

\smallskip
For the case where $l_{L,R}>l_C$, the tension bounds are given as~\cite{f,x}:
\begin{equation}\label{simplerbounds}
    \frac{1}{l_{C}} - \frac{1}{l_{L}} \leq 8\pi G \sigma_{1} \leq \frac{1}{l_{L}} + \frac{1}{l_{C}}\qquad  \qquad  \frac{1}{l_{C}} - \frac{1}{l_{R}} \leq 8\pi G \sigma_{2} \leq \frac{1}{l_{C}} + \frac{1}{l_{R}}
\end{equation}
As it turns out, for $l_{L,R}>l_C$ the only wedge possible extends on both sides of the interface line. But for other scenarios where one or both of $l_L$ and $l_R$ are smaller than $l_C$, the wedge may extend on both sides, or just one. In those scenarios, while imposing a central wedge that occurs on both sides of the interface permits $\sigma_1$ and $\sigma_2$ to be independently dialed in the solutions, it does constrain the lower bound to a higher value than the Coleman De Lucia bound in (\ref{simplerbounds}).

For $l_{L,R}<l_C$, while maintaining a central wedge on both sides of the interface line, the tension bounds are:
\begin{equation}\label{newbound}
    \sqrt{\frac{1}{l_{L}^2} - \frac{1}{l_{C}^2}} \leq 8\pi G \sigma_{1} \leq \frac{1}{l_{L}} + \frac{1}{l_{C}},\qquad  \qquad  \sqrt{\frac{1}{l_{R}^2} - \frac{1}{l_{C}^2}} \leq 8\pi G \sigma_{2} \leq \frac{1}{l_{C}} + \frac{1}{l_{R}}
\end{equation}
This new lower bound on tension also appears in~\cite{x} as a critical tension in a finite temperature AdS/CFT model\footnote{Below this value, the hot solution disappears for some part of the parameter space.}. 

For $l_L>l_C>l_R$, while maintaining a central wedge on both sides of the interface line, the tension bounds span:
\begin{equation}\label{eq:RCL}
    \frac{1}{l_{C}} - \frac{1}{l_{L}} \leq 8\pi G \sigma_{1} \leq \frac{1}{l_{C}} + \frac{1}{l_{L}},\qquad  \qquad  \sqrt{\frac{1}{l_{R}^2} - \frac{1}{l_{C}^2}} \leq 8\pi G \sigma_{2} \leq \frac{1}{l_{C}} + \frac{1}{l_{R}}
\end{equation}

Similarly, for $l_L<l_C<l_R$, while maintaining a central wedge on both sides of the interface line, the tension bounds span:
\begin{equation}\label{eq:LCR}
    \sqrt{\frac{1}{l_{L}^2} - \frac{1}{l_{C}^2}} \leq 8\pi G \sigma_{1} \leq \frac{1}{l_{L}} + \frac{1}{l_{C}} ,\qquad  \qquad  \frac{1}{l_{C}} - \frac{1}{l_{R}} \leq 8\pi G \sigma_{2} \leq \frac{1}{l_{C}} + \frac{1}{l_{R}}
\end{equation}

As we will see in Section 2.4, the lower tension bounds correspond to upper bounds on the transmission coefficient which is the same as the one following the achronal average-null-energy condition (AANEC). Therefore, this nuanced behaviour of the tension bounds and the position of the central wedge is noteworthy even though it isn't a feature directly observable from the boundary CFTs.

\smallskip
Our model has several parameters. In addition to the asymptotic curvature radii $l_L$ and $l_R$, which specify the central charges of the two CFTs connected by the interface, we have to specify several model parameters characterizing the interface itself: the central curvature radius $l_C$ and the string tensions $\sigma_1$ and $\sigma_2$. For the purposes of presenting our results, it is convenient to cut models down to a slightly smaller parameter space. One way to do this in a consistent manner is to restrict to systems which respect a $\mathbb{Z}_2$ parity symmetry across the interface. Obviously, this parity symmetric scenario is only possible if $l_L=l_R$, as well as $\sigma_1 = \sigma_2$. While more general forms of the results can be obtained, we will use the restricted parameter space when we are presenting our results in Figures \ref{fig3}, \ref{fig4}, \ref{fig5} and \ref{fig6}. The expressions for the general case are relegated to the appendix. For convenience we'll use $\tau_{1}$ and $\tau_{2}$ for the respective tensions with $\tau_{1} = 8\pi G \sigma_{1}$ and $\tau_{2} = 8\pi G \sigma_{2}$.

\smallskip
While in our drawing in Figure \ref{fig1} it appears that we are describing 4 branes, the right most brane of the left wedge is really the same brane as the left brane of the central wedge. These are one and the same object across which the various spacetime wedges get glued together. The same is true for the second apparent pair of branes. This identification between the branes translates to an identification of the metrics on the branes. In addition, we can use the Israel Junction conditions \cite{w} to assign the jump in the extrinsic curvature tensor across a brane to the tension of the corresponding string. Together these matching conditions are given by:
\begin{subequations}\label{eq:1}
\begin{align}
\label{eq:1:1}
\gamma_{L,\alpha \beta} &= \gamma_{CL,\alpha \beta} \\
\label{eq:1:2}
\gamma_{CR,\alpha \beta} &= \gamma_{R,\alpha \beta} \\
\label{eq:1:3}
K_{L,\alpha \beta}-K_{CL,\alpha \beta} - tr(K_{L}-K_{CL}) \gamma_{L, \alpha \beta} &= \tau_{1} \gamma_{L, \alpha \beta} \\
\label{eq:1:4}
K_{CR,\alpha \beta}-K_{R,\alpha \beta} - tr(K_{CR}-K_{R}) \gamma_{R, \alpha \beta} &= \tau_{2} \gamma_{R, \alpha \beta}
\end{align}
\end{subequations}
where $\gamma_{L,CL,CR,R}$ and $K_{L,CL,CR,R \alpha \beta}$ are the induced metric and extrinsic curvature tensor on the branes suspended by $M_{L,C,R}$ respectively.

\smallskip
In pure AdS$_{3}$, Einstein's equations can be solved completely and the full solution for the metric in Fefferman-Graham co-ordinates can be written as~\cite{k}:
\begin{equation} 
\label{26}
    ds^{2} = \frac{l^{2}dy^{2}}{y^{2}} + \left[ \frac{l^{2}}{y^{2}}g_{\alpha \beta}^{(0)} + g_{\alpha \beta}^{(2)} + \frac{y^{2}}{4l^{2}}g_{\alpha \beta}^{(4)} \right] dw^{\alpha}dw^{\beta}
\end{equation}
with $g^{(4)} = g^{(2)} (g^{(0)})^{-1} g^{(2)}$. For a flat boundary metric, we can furthermore identify $g_{\alpha \beta}^{(2)} = 4Gl<T_{\alpha \beta}>$ with $<T_{\alpha \beta}>$ being the vacuum expectation value of the canonically-normalized, traceless conserved energy-momentum tensor in some state of the dual CFT. 

\smallskip
Therefore the ICFT vacuum metric in Fefferman-Graham co-ordinates for the three regions is given by~\cite{d}:
\begin{equation}
\label{eq:x}
\begin{split}
ds_{L}^{2} &= \frac{l_{L}^2}{y_{L}^{2}}\left[ dy_{L}^{2} +  du_{L}^{2} - dt_{L}^{2} \right]  \qquad u_{L} \leq y_{L}\tan{\theta_{L}}\\
ds_{R}^{2} &= \frac{l_{R}^2}{y_{R}^{2}}\left[ dy_{R}^{2} +  du_{R}^{2} - dt_{R}^{2} \right] \qquad u_{R} \geq y_{R} \tan{\theta_{R}} \\
ds_{C}^{2} &= \frac{l_{C}^2}{y_{C}^{2}}\left[ dy_{C}^{2} +  du_{C}^{2} - dt_{C}^{2} \right] \qquad y_{C} \tan{\theta_{CL}} \leq u_{C} \leq y_{C} \tan{\theta_{CR}}
\end{split}
\end{equation} 
where $0<y_{L,R,C}<\infty$ and the branes subtend at angles where the inequalities saturate.

\smallskip
In general, all the angles $\theta_{L,R,CL,CR}$ can be positive or negative as long as they satisfy $(\theta_{CL}-\theta_{CR})>0$ which imposes that some non-zero central region exists. For the $M_C$ wedge extending on both sides of the interface, $\theta_{CL}<0$ and $\theta_{CR}>0$ while $\theta_{L}$ and $\theta_{R}$ can be positive or negative, subject to the AdS radii and tension bounds.

\smallskip
Like the setup in~\cite{y,f}, we consider right-moving monochromatic gravitational wave excitations coming from the left side on the surface. We solve the matching equations in \eqref{eq:1} up to linear order in $\epsilon$, the magnitude of the incoming flux, following the technique in~\cite{f}. This allows us to drop the $g^{(4)}$ term from the full metric solution and the correction to the AdS$_{3}$ Poincare metric just has arbitrary left and right moving waves $g^{(2)}_{++}(w^{+})$ and $g^{(2)}_{--}(w^{-})$. Here we can identify $w^{\pm}$ as $u \pm t$. 

\smallskip
The incoming surface monochromatic wave from the left undergoes reflections and transmissions at both branes, leading to a reflected and transmitted wave on the surface of $M_L$ and $M_R$ respectively, with a left and right moving wave in $M_C$. Incoming excitation corresponds to $\left<T_{--}\right>$ and all subsequent reflections/transmissions are expressed in terms of this, the reflection/transmission coefficient from the corresponding boundary and the corresponding left/right moving exponential. The corrections to the AdS$_{3}$ Poincare metrics in the three regions are given as: 
\begin{equation}
\begin{split}
    [{d s}^{2}_{L}]^{(2)} &= 4 G l_{L} \epsilon \left[ e^{i w(t_{L}-u_{L})}d(t_{L}-u_{L})^{2} + \mathcal{R}_{L1}e^{i w(t_{L}+u_{L})}d(t_{L}+u_{L})^{2} \right] + c.c. \\
    [{d s}^{2}_{C}]^{(2)} &= 4 G l_{C} \epsilon \left[ \mathcal{T}_{L1}e^{i w(t_{C}-u_{C})}d(t_{C}-u_{C})^{2} + \mathcal{R}_{L2}e^{i w(t_{C}+u_{C})}d(t_{C}+u_{C})^{2} \right] + c.c. \\
    [ds^{2}_{R}]^{(2)} &= 4 G l_{R} \epsilon \left[ \mathcal{T}_{L2}e^{i w(t_{R}-u_{R})}d(t_{R}-u_{R})^{2} \right] + c.c.
\end{split}
\end{equation}
where $|\left<T_{--}\right>|= \epsilon$ is the magnitude of the incoming flux and $\mathcal{R}_{L1}, \mathcal{T}_{L1}, \mathcal{R}_{L2}$ and $\mathcal{T}_{L2}$ are the a priori complex relative amplitudes of the reflected and transmitted waves from the first and second brane respectively. The subscript L indicates the incident wave originated on the left hand side.

\smallskip
In order to glue the metrics on the branes, it will first be easier to rotate from the Poincare co-ordinates to those parallel(z) and perpendicular(x) to each brane. This rotation is given by: 
\begin{equation}
        \begin{pmatrix} u_{L} \\ y_{L} \end{pmatrix}
        = 
        \begin{pmatrix} \cos{\theta_{L}} & \sin{\theta_{L}} \\ - \sin{\theta_{L}} & \cos{\theta_{L}} \\ \end{pmatrix}
        \begin{pmatrix} x_{L} \\ z_{L} \end{pmatrix}
\end{equation}
$(x_{R},z_{R}),(x_{CL},z_{CL}),(x_{CR},z_{CR})$ transform similarly using their respective angles $\theta_{R}$, $\theta_{CL}$ and $\theta_{CR}$.

\smallskip
Gluing one pair of branes together requires matching co-ordinates on the worldsheet that gives the AdS$_2$ vacuum metric. Additionally, due to time translation invariance, the co-ordinates can be defined for a fixed frequency $w$. Such co-ordinates can be expressed as: 
\begin{equation}\label{eq:functions}
    \begin{split}
        t_{L} &= t_{1} + \Tilde{\epsilon} e^{iwt_{1}}\lambda_{L}[z_{1}],  \\
        t_{R} &= t_{2} + \Tilde{\epsilon} e^{iwt_{2}}\lambda_{L}[z_{2}],  \\
        t_{CL} &= t_{1} + \Tilde{\epsilon} e^{iwt_{1}}\lambda_{CL}[z_{1}], \\
        t_{CR} &= t_{2} + \Tilde{\epsilon} e^{iwt_{2}}\lambda_{CR}[z_{2}], \\
    \end{split}
    \qquad
    \begin{split}
        z_{L} &= z_{1} + \Tilde{\epsilon} e^{iwt_{1}}\zeta_{L}[z_{1}], \\
        z_{R} &= z_{2} + \Tilde{\epsilon} e^{iwt_{2}}\zeta_{R}[z_{2}], \\
        z_{CL} &= z_{1} + \Tilde{\epsilon} e^{iwt_{1}}\zeta_{CL}[z_{1}],\\
        z_{CR} &= z_{2} + \Tilde{\epsilon} e^{iwt_{2}}\zeta_{CR}[z_{2}],
    \end{split}
    \qquad
    \begin{split}
        x_{L} &= \Tilde{\epsilon} e^{iwt_{1}}\delta_{L}[z_{1}]. \\
        x_{R} &= \Tilde{\epsilon} e^{iwt_{2}}\delta_{R}[z_{2}]. \\
        x_{CL} &= \Tilde{\epsilon} e^{iwt_{1}}\delta_{CL}[z_{1}]. \\
        x_{CR} &= \Tilde{\epsilon} e^{iwt_{2}}\delta_{CR}[z_{2}].
    \end{split}
\end{equation}
where $(t_1,z_1)$ and $(t_2,z_2)$ are the Poincare cooridnates on the AdS$_{2}$ metrics for the first and second brane respectively. $\Tilde{\epsilon} = \frac{4G\epsilon}{l_{S1,S2}}$ is used for convenience.

\smallskip
Applying the matching conditions to the ICFT vacuum (0 order in $\epsilon$) for the induced metric shows that the worldsheet metric is AdS$_{2}$ with radius $l_{S1}$ and $l_{S2}$ for the left and right tensile strings respectively and gives the first two equalities in \eqref{eq:4}. The jump in the extrinsic curvature condition gives the last equality: 
\begin{equation}\label{eq:4}
\begin{split}
    l_{S1} &= \frac{l_{L}}{\cos{\theta_{L}}} =\frac{l_{CL}}{\cos{\theta_{CL}}} = \frac{\tan{\theta_{L}} - \tan{\theta_{CL}}}{\tau_{1}}\\
     l_{S2} &= \frac{l_{CR}}{\cos{\theta_{CR}}} =\frac{l_{R}}{\cos{\theta_{R}}} = \frac{\tan{\theta_{CR}} - \tan{\theta_{R}}}{\tau_{2}}
\end{split}
\end{equation}

\smallskip
Applying matching conditions of order 1 in $\epsilon$ leads to a set of linearized differential equations instead of simplified constraints as above. The differential equations and their solutions are presented in Appendix \ref{appa}. Below, we will only highlight that instead of the 12 unknown functions we've used to describe the co-ordinates, what actually appears in the differential equations are re-parametrizations of the above functions, some of which represent physical sources that can fluctuate the interface in different ways. 

\smallskip
There are 12 unknown functions and 12 matching equations (symmetric metric and tensors). However, the extrinsic curvature tensor's elements are not all independent as they obey a momentum constraint. Thus, Israel junction condition gives 1 equation instead of 3. That leaves a total of 8 constraints and 12 functions. Due to a re-parametrization, only the following functions actually appear: 
\begin{equation}
    \begin{split}
        \lambda_{1} &= \lambda_{L}-\lambda_{CL}, \qquad \zeta_{1}= \zeta_{L} - \zeta_{CL}, \\
        \lambda_{2} &= \lambda_{CR}-\lambda_{R}, \qquad \zeta_{2}= \zeta_{CR} - \zeta_{R}. 
    \end{split}
\end{equation}
Thus, there are 8 constraints and 8 functions. Furthermore, the matching equations are expressed neatly using the following definitions: 
\begin{equation}
\begin{split}
    D_{1} &= \delta_{L}-\delta_{CL}, \qquad \Delta_{1} = \tan{\theta_{L}} \delta_{L} - \tan{\theta_{CL}} \delta_{CL} - \zeta_{1}, \\
    D_{2} &= \delta_{CR}-\delta_{R}, \qquad \Delta_{2} = \tan{\theta_{CR}} \delta_{CR} - \tan{\theta_{R}} \delta_{R} - \zeta_{2}. 
    \end{split}
\end{equation}
The $z_1 = 0$ and $z_2 = 0$ limits of these functions correspond to sources in the dual ICFT - fluctuating the interface. D$_1(0)$ and D$_2(0)$ are sources for the interface displacement operator and $\lambda_1(0)$ and $\lambda_2(0)$ for an operator reparametrizing the interface. 

\smallskip
Appendix \ref{appa} can show that only the re-parametrizations stated above appear as variables in the differential equations. You can solve for the homogeneous solution to the differential equations, that is where no gravitational waves are scattered at the boundary. The homogeneous solutions for $\lambda_1, \zeta_1$ and $\lambda_2, \zeta_2$ all comprise of linear combinations of solutions going in and out of the horizon (travelling along the branes). The coefficients for those modes are labelled $a_{1-}, a_{1+}$ and $a_{2-}, a_{2+}$. The solution for D$_1$ and D$_2$ are constants. 

\smallskip
However, for a non-fluctuating interface such as the one we have, we need to set the sources to zero. For example $D_1(0) = 0 \rightarrow \delta_{L}(0) = \delta_{CL}(0)$ so there is no relative displacement of the interface. For the homogeneous solutions, this sets the constants for D$_1$ and D$_2$ to 0. Imposing the other sources to 0 at the boundary ends up setting all the incoming and outgoing solutions on both the right and left branes to 0. Therefore, no solutions in the empty background can be supported by a purely non-fluctuating interface.

\subsection{Results and comments}
We can solve for the particular solution to the full differential equations, which is present in Appendix \ref{appa}. Considering the non-fluctuating interface, we set all the sources to go to 0 at the boundary.

$D_1(0)=0$ and $D_2(0)=0$ gives: 
\begin{equation}
    \mathcal{R}_{L1} + \mathcal{T}_{L2} = 1
\end{equation}

$\lambda_1(0)=0$, $\zeta_1(0)=0$, and $\lambda_2(0)=0$, $\zeta_2(0)=0$ give expressions for the modes going in and out of the horizon: $a_{1-}, a_{1+}, a_{2-}, a_{2+}$ (travelling along the branes in both directions, for both branes). These expressions can be found in Appendix \ref{appa}.  

\smallskip 
We now impose the physical condition that both the modes coming out of the horizon, $a_{1+}$ and $a_{2+}$, should be removed as they are not physical~\cite{z,aa}. So $D_1(0)=0$, $D_2(0)=0$, $a_{1+}=0$ and $a_{2+}=0$ give expressions for the transmission/reflection coefficients. The expressions are in terms of the angles, and can be found in Appendix \ref{appa}. Simplifying them using the relations from \eqref{eq:4}, we obtain: 
\begin{equation}
\label{215}
  \mathcal{T}_{L2} = \frac{2}{l_{L}} \left( \frac{1}{l_{L}} + \frac{1}{l_{R}} + \tau_1 + \tau_2 \right) ^{-1}
\end{equation}
Note that the AdS radius of the central region plays an important role in constraining the bounds of the tensions but does not explicitly appear in the expression for transmission or reflection coefficient on either side of the physical CFT regions with non-zero length. In particular, to take the limit of $l_C \rightarrow 0$ the tensions $\tau_{1,2}$ have to go to infinity and the transmission vanishes. This allows a manifest bulk realization of the limit of two decoupled BCFTs: there is simply no space left in the middle wedge to connect the two BCFTs.

\smallskip
The transmission and reflection coefficient for the waves that would live within the central region is given by: 
\begin{equation}
    \begin{split}
    \mathcal{R}_{L2} &= 
    \frac{l_C}{2} \left( \frac{1}{l_{R}} - \frac{1}{l_{C}} + \tau_2 \right) \mathcal{T}_{L2} \\
    \mathcal{T}_{L1} &= 
    \frac{l_C}{2} \left( \frac{1}{l_{R}} + \frac{1}{l_{C}} + \tau_2 \right) \mathcal{T}_{L2} 
    \end{split}
\end{equation}

\smallskip
To highlight the difference in how the transmission coefficient depends on the AdS radius, consider the transmission coefficient for the single tensile string case~\cite{f}: 
\begin{equation}\label{eq:oldT}
  \mathcal{T}_{L} = \frac{2}{l_{L}} \left( \frac{1}{l_{L}} + \frac{1}{l_{R}} + \tau_1 \right) ^{-1}
\end{equation}

\begin{figure}[!ht]
\centering % \begin{center}/\end{center} takes some additional vertical space
\includegraphics[width=.75 \textwidth]{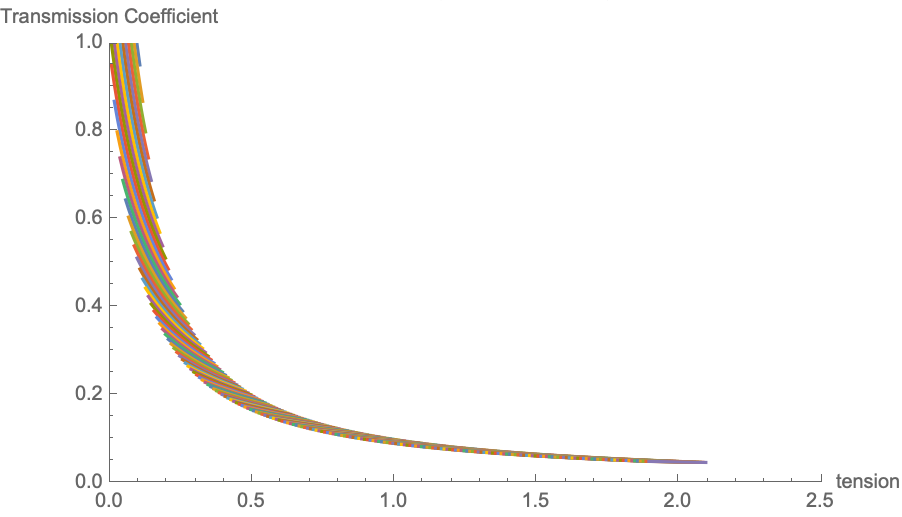}
% "\includegraphics" is very powerful; the graphicx package is already loaded
\caption{\label{fig2} Transmission coefficient for the 2-region holographic model for different $l_L$ and $l_R$ where $l_L/l_R$ ranges from $0.05$ to $20$.}
\end{figure}

\begin{figure}[!ht]
\centering % \begin{center}/\end{center} takes some additional vertical space
\includegraphics[width=.75 \textwidth]{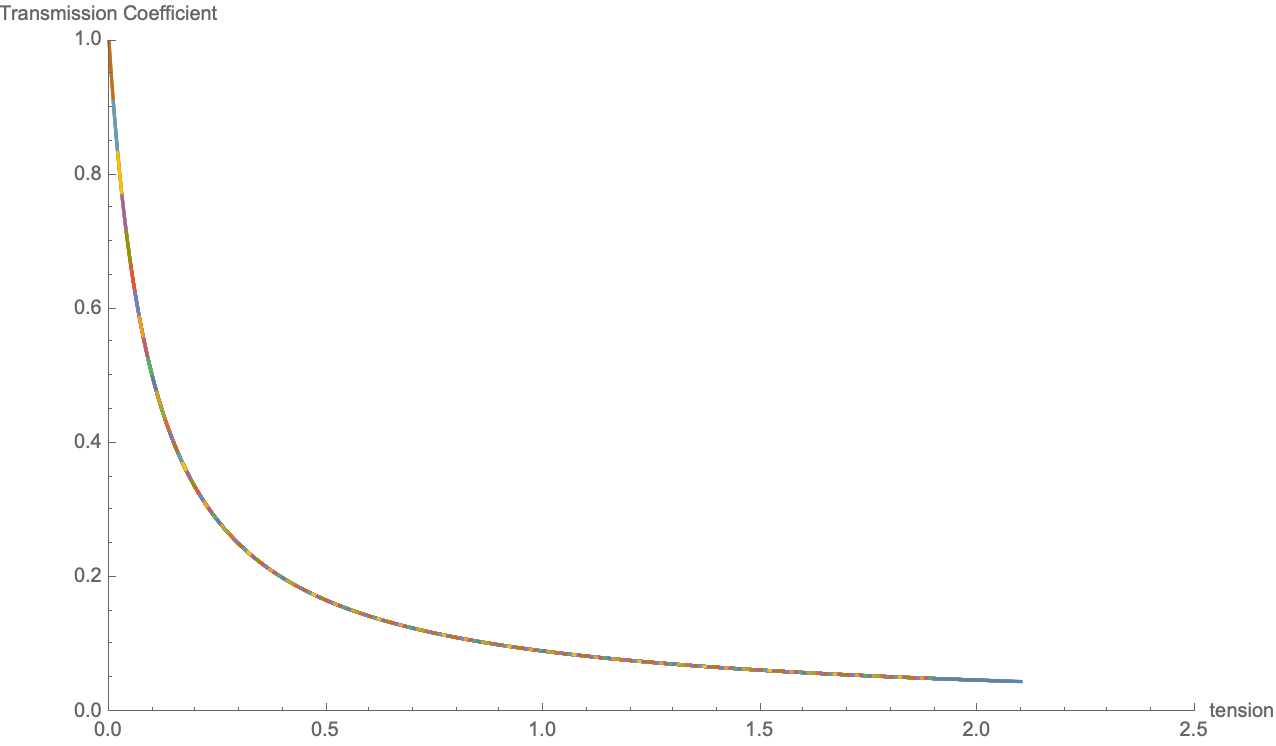}
% "\includegraphics" is very powerful; the graphicx package is already loaded
\caption{\label{fig3} Transmission coefficient for the 3-region holographic model where we consider a symmetric fusion of defects $l_L=l_R$ and $\tau_1=\tau_2$, where $l_L/l_C$ ranges from $0.05$ to $20$.}
\end{figure}

$\mathcal{T}_{L2}$ in \eqref{eq:oldT} and \eqref{215} has the same form as a function of $\tau_1$ and the net tension ($\tau_1+\tau_2$) respectively but it is important to keep in mind that the 3-region model has additional freedom in $l_C$. 
Figure \ref{fig2} shows the different transmission coefficient curves for the 2-region model \eqref{eq:oldT} for various combinations of $l_L$ and $l_R$. Figure \ref{fig3} shows the transmission coefficient \eqref{215} for the parity symmetric 3-region model ($l_L =l_R$ and $\tau_1 = \tau_2$) for different $l_C$ values. For a given $l_L$ and $l_R$, figure \ref{fig2} gives a single transmission coefficient curve dialled by tension with a fixed beginning and end point, while in figure \ref{fig3}, different $l_C$ values correspond to segments\footnote{Segments may overlap.} (transmission curves dialled by the tension) on an overall curve. For a given $l_L$ and $l_R$, fixing the tension (the only parameter) in the 2-region model gives exactly one point on the transmission curve. The important difference in the parity symmetric 3-region model is that we can instead fix $l_C$ and cover a finite range of transmission values by dialling the other parameter, tension. Here we fix $l_C$ but alternatively we could choose to fix $g$ instead.\footnote{(In the interest of completeness, let us say in advance) For a fixed $l_{L,R}$ and $\log \ g$, the single brane corresponds to one tension and therefore one point on the transmission curve, but for the double brane, it correlates to multiple possible ($l_C$ values and so) tensions and therefore multiple points on the transmission coefficient curve.} Either way, a range of transmission coefficients are possible.

Besides the difference in the way in which the transmission coefficient depends on the boundary's radius and the center's radius, the other difference is in the range of transmission coefficient values as one goes from the 2-region to the 3-region model. The lower and upper bounds for the transmission coefficient can be obtained by using the upper and lower limits of tensions respectively. $\mathcal{T}_{L2}$'s upper bound would differ depending on how the AdS radii vary.
\begin{equation} \label{NewTransmissionBounds}      
    \begin{split}
        \mathcal{T}_{L2_{min, all}} &= \left( 1 + \frac{l_L}{l_R} + \frac{l_L}{l_C} \right)^{-1} \\
        \mathcal{T}_{L2_{max,A}} &= \frac{l_C}{l_L}  \\
        \mathcal{T}_{L2_{max,B}} &= \frac{2}{1+\frac{l_L}{l_R} + \sqrt{1-(\frac{l_L}{l_C})^2} + \sqrt{(\frac{l_L}{l_R})^2-(\frac{l_L}{l_C})^2}}
    \end{split}
    \begin{split}
        \qquad \\
        \mathcal{T}_{L2_{max,C}} &= \frac{2}{ \frac{l_L}{l_R} + \frac{l_L}{l_C} + \sqrt{(\frac{l_L}{l_R})^2-(\frac{l_L}{l_C})^2} }\\
        \ \ \mathcal{T}_{L2_{max,D}} &= \frac{2}{1+\frac{l_L}{l_C} + \sqrt{1-(\frac{l_L}{l_C})^2}} \\
    \end{split}
\end{equation}
where A, B, C and D respectively denote scenarios $l_{L,R}>l_C$, $l_{L,R}<l_C$, $l_L>l_C>l_R$ and $l_L<l_C<l_R$.

In the single brane case, the upper bound on the transmission coefficient \cite{f} is the same as the one following from the achronal average-null-energy condition (AANEC)~\cite{y}. In the double brane model, we observe the maximum transmission coefficient values are all consistent with the AANEC bound, and for specific $l_C$ values saturate it for scenarios A, C and D, but not for B (in general).

The AANEC bound is given by \cite{y}:
\begin{equation}\label{eq:AANEC}
\begin{split}
    \mathcal{T}_{L2} &\leq \frac{l_R}{l_L} \qquad l_R<l_L\\ 
    \mathcal{T}_{L2} &\leq 1 \qquad \quad l_R>l_L
\end{split}
\end{equation}

When $l_{L,R}>l_C$ (A), (\ref{NewTransmissionBounds}) shows $\mathcal{T}_{L2_{max}}$ is consistent with (\ref{eq:AANEC}) for all allowed values of $l_C$, and saturates the AANEC bound as $l_C = l_R$ for $l_R<l_L$ and $l_C = l_L$ for $l_R>l_L$. 

\smallskip
In the regimes where $l_C$ is intermediate (C and D), $\mathcal{T}_{L2_{max}}$ is continuous and saturates the AANEC bound in \ref{eq:AANEC} for limits on $l_C$. For $l_L>l_C>l_R$ (C), \ref{NewTransmissionBounds} saturates the bound in the limit $l_C \xrightarrow{} l_{R}$ from above. For $l_L<l_C<l_R$ (D), it saturates the bound in the limit $l_C \xrightarrow{} l_{L}$ from above.

\smallskip
When $l_{L,R}<l_C$ (B), $\mathcal{T}_{L2_{max}}$ has no local maxima as a function of $l_C$, so the maximum value is reached at either $l_C$'s maximum value of $\infty$, or at $l_C$'s minimum value, given by maximum($l_L,l_R$). From (\ref{NewTransmissionBounds}), as $l_C \xrightarrow{} \infty, \mathcal{T}_{L2_{max}} = \frac{l_R}{l_R+l_L}$. While this is always consistent with \ref{eq:AANEC}, it only saturates the AANEC bound in the (trivial) limiting scenario: $l_R \to 0$ for $l_R<l_L$, and $l_L \to 0$ for $l_R>l_L$. As $l_C$ approaches it's minimum allowed value: 
\begin{equation}
\begin{split}
    \text{For } l_R<l_L \qquad l_C \xrightarrow{} l_L \qquad \mathcal{T}_{L2_{max}} &= \frac{2}{1+\frac{l_L}{l_R} +  \sqrt{(\frac{l_L}{l_R})^2-1}} < \frac{l_R}{l_L}\\
    \text{For } l_R>l_L \qquad l_C \xrightarrow{} l_R \qquad \mathcal{T}_{L2_{max}} &= \frac{2}{1+\frac{l_L}{l_R} + \sqrt{1-(\frac{l_L}{l_R})^2} } < 1
\end{split}
\end{equation}
This does not saturate the AANEC bound in general. It is  consistent with $\ref{eq:AANEC}$ but only saturates for the specific case $l_R=l_L$ (or in the (trivial) limiting scenario mentioned above).

\begin{figure}[!ht]
\centering % \begin{center}/\end{center} takes some additional vertical space
\includegraphics[width=.75 \textwidth]{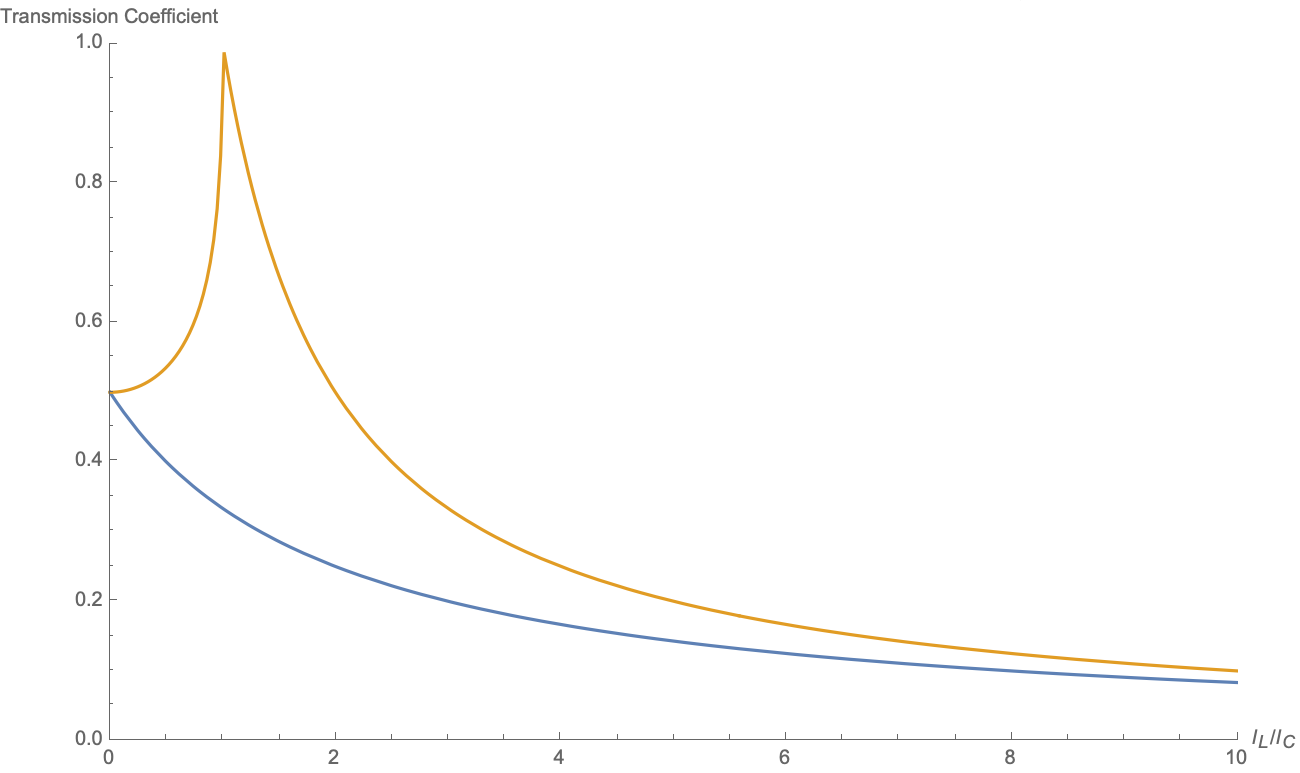}
% "\includegraphics" is very powerful; the graphicx package is already loaded
\caption{\label{fig4} The upper and lower bound of the transmission coefficient for the 3-region holographic model with the same boundary radius. As $l_c \rightarrow 0$ both upper and lower bound go to 0.}
\end{figure}

\smallskip
We look at the upper and lower bounds of the transmission coefficient for the specific case where the boundary radii are the same. The result is shown in Figure \ref{fig4}. We can see that the maximum difference between the upper and lower bound of the transmission coefficient occurs for the limiting case where all three radii are the same. The values of $\mathcal{T}_{L2}$ for $l_L=l_R=l_C$ can be given by
\begin{equation}
   \mathcal{T}_{L2_{min}} = \frac{1}{3}, \qquad  \mathcal{T}_{L2_{max}}=1. 
\end{equation}

This is in contrast with the 2-region model where the lower bound is $1/2$ for the case where both the boundary radii are the same. This goes to show that for the same AdS radii, one can dial the transmission coefficient's minimum value down by adding central wedges that represent point-like defects lying on top of one another. As the number of wedges increases, the minimum transmission coefficient goes to 0.

\section{Calculating entanglement entropy to find g}

In holography, we know that entanglement entropy for a subsystem M, of length $a$, is given by the area of the minimum extremal RT surface in the d$+1$ dimensional bulk suspended by the boundary of the d-dimensional CFT subsystem M:~\cite{ab,ac}
\begin{equation*}
S = \frac{Area_{min}}{4G^{d+1}_N}
\end{equation*}
where G$^{d+1}_N$ is the d$+1$ dimensional Newton's constant.

\smallskip
For an AdS$_3$ background, this minimal area turns out to be simpler in global co-ordinates. Therefore, it will be advantageous to express the metric \eqref{eq:x} and the constraints from the matching equations \eqref{eq:4} in these. 
\subsection{Global co-ordinates}
Relating the Poincare coordinates $(t, u, y)$ to the global co-ordinates $(t, x, r)$\footnote{x is a radial co-ordinate here, not the co-ordinate orthogonal to the branes as in Section 2.}:
\begin{equation}
\label{coc}
    y_{L} = \frac{x_{L}}{\cosh{\left( \frac{r_{L}}{l_{L}} \right)} } \qquad u_{L} = x_{L} \tanh{\left( \frac{r_{L}}{l_{L}} \right)}
\end{equation}
The AdS$_{3}$ metric in global co-ordinates is given by: 
\begin{equation}
    \begin{split}
        ds^{2}_{L} &= l_{L}^2 \cosh^2 \left( \frac{r_L}{l_{L}}  \right) ds^2_{2} + dr_L^2 \qquad -\infty <r_L< R_L  \\
        ds^{2}_{C} &= l_{C}^2 \cosh^2 \left( \frac{r_C}{l_{C}} \right) ds^2_{2} + dr_C^2 \qquad R_{CL} <r_C< R_{CR} \\
        ds^{2}_{R} &= l_{R}^2 \cosh^2 \left( \frac{r_R}{l_{R}}\right) ds^2_{2} + dr_R^2 \qquad R_R <r_R< \infty 
    \end{split}
\end{equation}
where $r_L$,$r_C$, $r_R$ are the hyperbolic angles and $ds_{2}^{2}$ is the metric for a constant $r$ AdS$_{2}$ slice given by: 
\begin{equation}
    ds^{2}_{2} = \frac{-dt_L^{2}+ dx_L^{2}}{x_L^{2}} = -\cosh^2{\mu} \ dt^{2} + d\mu^{2}
\end{equation}
$ds^{2}_{2}$ can similarly be expressed in (t$_C$, x$_C$) and (t$_R$, x$_R$). For the same reasons stated in Section 2, note that $R_L$, $R_{CL}$, $R_{CR}$ and $R_R$ can in general be positive or negative as long as $(R_{CR}- R_{CL})>0$. An $M_C$ wedge extending on both sides of the interface line translates to $R_{CL}<0$ and  $R_{CR}>0$.

\smallskip
The matching conditions \eqref{eq:1} for the vacuum solution are given as:
\begin{equation}
\label{34}
    \begin{split}
        l_{L}\cosh{\left( \frac{R_{L}}{l_{L}} \right) } &= l_{C}\cosh{\left( \frac{R_{CL}}{l_{C}} \right)} \\
        \frac{Tanh \left( \frac{R_{L}}{l_{L}} \right)}{l_{L}}  &- \frac{Tanh \left( \frac{R_{CL}}{l_{C}} \right)}{l_{C}} = \tau_{1}
    \end{split}
    \qquad
    \begin{split}
        l_{C}\cosh{\left( \frac{R_{CR}}{l_{C}} \right)} &= l_{R}\cosh{\left( \frac{R_{R}}{l_{R}} \right)} \\
        \frac{Tanh \left( \frac{R_{CR}}{l_{C}} \right)}{l_{C}}  &- \frac{Tanh \left( \frac{R_{R}}{l_{R}} \right)}{l_{R}}  = \tau_{2}
    \end{split}
\end{equation}

\subsection{Review: Entanglement entropy for a 2-region ICFT}
Entanglement entropy for a subsystem that encloses a defect has been worked out in~\cite{n,g}. For a symmetric interval of total length a, that is where the defect is at the center of the interval, the entanglement entropy of this subsystem is given as:
\begin{equation}\label{oldeeform}
S = \frac{(l_{R} + l_{L})Log(\frac{a}{\epsilon})}{4G} + \log g_{0}
\end{equation}
where $\epsilon$ is the UV cut-off, and G is the 3 dimensional Newton's constant. 

For the single brane model, the presence of the defect provides the additional contribution to the entanglement entropy given by: 
\begin{equation}\label{oldLogG}
    \log g_{0} = \frac{1}{4G}(R_{L} - R_{R})
\end{equation}

\subsubsection{Entanglement entropy for a 3-region ICFT}
The method from~\cite{g} is implemented for the 3-region holographic model to calculate the area of the minimal RT surface and the entanglement entropy form. 

To write the entanglement entropy, we start with the minimal area functional - as this method is the same for all three regions, for this part of the calculation, we will drop the subscript L from the AdS radius and from the co-ordinates $(t,r,x)$. This is given by the square root of the metric, for a constant time slice ($t=0$) and parameterized by $r$ so that $x$ becomes $x(r)$: 
\begin{equation}
\label{37}
    \mathcal{A} = \sqrt{l^2 \cosh^2 \left( {\frac{r}{l}}\right)\frac{x'^2}{x^2} + 1} \ \ dr
\end{equation}
where $x'$ is $\frac{\partial x}{\partial r}$. 

The constraint on $x(r)$ is obtained by using the scale isometry of AdS$_2$, that is $x \rightarrow \lambda x$ is a symmetry of $\mathcal{A}$ that corresponds to a Noether charge. Noether's charge for some functional $\mathcal{L}$ with fields $\psi_i$ is given below and when applied to \eqref{37} gives:
\begin{equation}
    \begin{split}
        j^{\mu} &= -\mathcal{L}\delta x^\mu + \frac{\partial \mathcal{L}}{\partial (\partial_\mu \psi_i)} (\partial_\alpha \psi_i)\delta x^\alpha \\
        c_s &= \frac{l^2 \cosh^2{(\frac{r}{l})}x'}{\sqrt{l^2\cosh^2{(\frac{r}{l})}x'^2 + x^2}}
    \end{split}
    \label{38}
\end{equation}
Solving for $\frac{x'}{x}$ and plugging into Eq \eqref{37}, we get:
\begin{equation}
    \mathcal{A} = \frac{l \cosh(\frac{r}{l})}{\sqrt{l^2 \cosh^2{(\frac{r}{l})}- c_s^2}} dr
\end{equation}
From \eqref{38}, $c_s=0$ corresponds to $x' =0$ so $x(r)=$ constant. To fix this constant, we look independently at our regions. For $r= \infty$, $x = a_R$ (area curve hits the right boundary of the subsystem) and for $r= - \infty$, $x = a_L$ (area curve hits the left boundary of the subsystem). In order to have the minimal area surface continuous across the brane, $a_L =  a_R = a/2$. Therefore, $c_s=0$ corresponds to the symmetric case where the defect is in the center of the interval. Then, the minimal area is given by: 
\begin{equation}
\label{310}
    \mathcal{A} = \int dr = \int^{R_L}_{-\infty} dr + \int^{R_{CR}}_{R_{CL}} dr + \int^{\infty}_{R_R}  dr
\end{equation}
This area needs to be regulated by truncating the integral as it goes to large positive and negative $r$'s $(r_+$ and -$r_{-})$. Using \eqref{coc} and setting $y = \epsilon$ as the UV cut-off, and $x_L =  x_R = a/2$:
\begin{equation}
\label{311}
    \begin{split}
        \frac{e^{\frac{r_+}{l_R}}}{2} &= \frac{x_R}{y_R} \qquad \qquad \qquad \frac{e^{\frac{r_-}{l_L}}}{2} = \frac{x_L}{y_L} \\
        r_+ &= \log (\frac{a}{\epsilon})l_R \qquad  \qquad r_- = \log (\frac{a}{\epsilon})l_L
    \end{split}
\end{equation}
Using \eqref{310} and \eqref{311} gives the entanglement entropy for a symmetric, defect enclosing, interval of length $a$: 
\begin{equation}\label{neweeform}
S = \frac{(l_{R} + l_{L})Log(\frac{a}{\epsilon})}{4G} + \log g_{1}
\end{equation}
\begin{equation}
\label{313}
    \log g_{1} = \frac{1}{4G}(R_{L} - R_{CL} + R_{CR}- R_{R})
\end{equation}

\smallskip
In general, $\log\ g$ is additive under the fusion of branes, evidenced by how (\ref{oldLogG}) extends to (\ref{313}). The constraints from the matching conditions \eqref{34} can be used to calculate these 4 unknowns at which the branes end and then using \eqref{313}, express $\log g_{1}$. Since these expressions don't simplify out in general, we lay out their details in Appendix \ref{appb}. 

\smallskip
\begin{figure}[!ht]
\centering
\includegraphics[width=1 \textwidth]{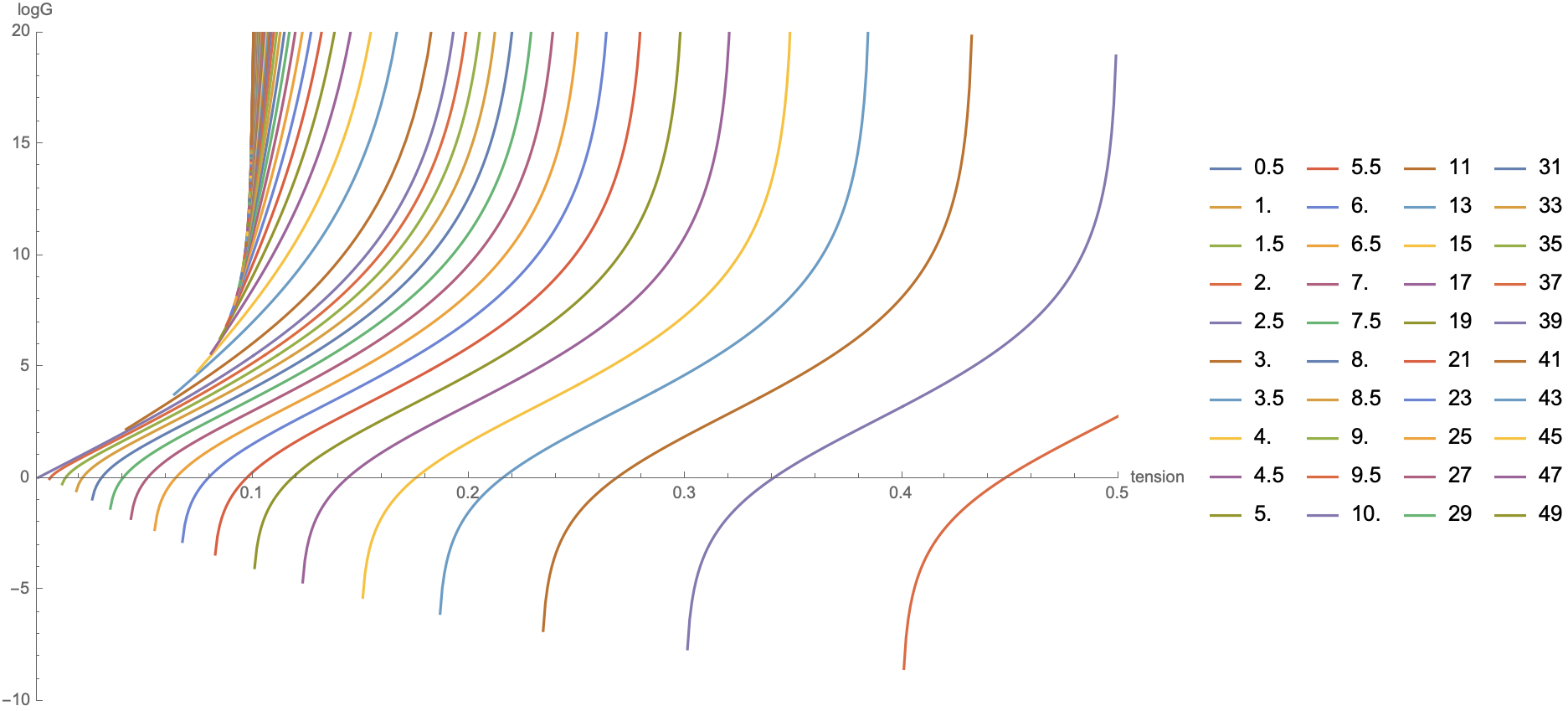}
\caption{\label{fig5} The additional entanglement entropy ($\log g$) for the 3-region holographic model, where we consider a symmetric fusion of defects with $l_L=l_R=10$, $\tau_1=\tau_2$. Legend shows $l_C$ values. Moving left to right, the curves are given by decreasing $l_C$ values.}
\end{figure}
\subsection{Results and comments}
\begin{figure}[!ht]
\centering 
\includegraphics[width=.50 \textwidth]{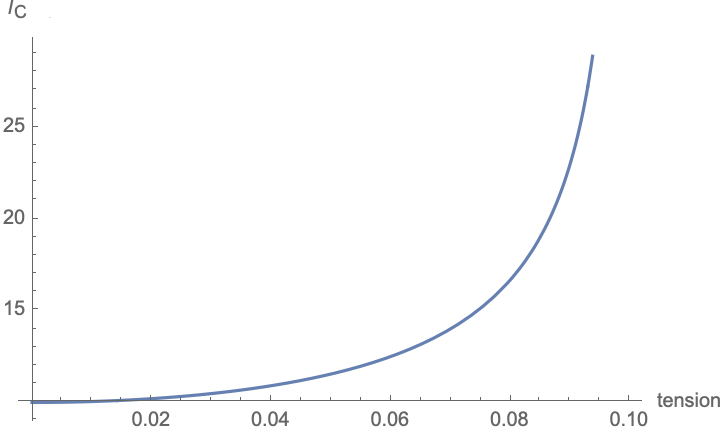}
\caption{\label{fig6} $\log g$ has a finite maximum value when $l_C>l_L (l_L=l_R=10$ and $\tau_1=\tau_2)$.}
\end{figure}

It is more expository to instead plot $\log g_{1}$. For an easy to visualize 2d plot, we take $l_L=l_R$ and $\tau_1=\tau_2$ representing a parity symmetric 'fusion' of defects. Figure \ref{fig5} shows that the correction to entanglement entropy can be positive or negative and unbounded in general, for different values of tension. Most interestingly, it shows that for the same $\log g_{1}$ value, we can have many different 3-region models, given by different $l_C$. This is exactly the freedom we were looking for that was absent in the single brane model. Additionally, we see that when the center's radius is greater than boundary's radius, there seems to be low tension region where $\log g_{1}$ has a maximal upper bound, depending on $l_C$. This can be explicitly seen in Figure \ref{fig6}. 

Using \eqref{logGbig} in Appendix \ref{appb}, if one evaluates $\log g_{1}$ for $l_{R}=l_{L}=l_{C}$, and then converts from log to $\tanh^{-1}$ (using the mathematical identity), the increment in entropy is:  
\begin{equation}
\label{314}
    \begin{split}
    \Delta S &= \frac{l_L \log \left(\frac{\left(\tau _1 l_L+2\right) \left(\tau _2 l_L+2\right)}{\left(\tau _1 l_L-2\right) \left(\tau _2 l_L-2\right)}\right)}{4 G} \\
    &= \frac{c_L}{3} \left( \tanh^{-1}{\left( \frac{l_L \tau_{1}}{2}\right)} + \tanh^{-1}{\left( \frac{l_L\tau_{2}}{2}\right)} \right)\\
    &= \log g_{0_1} +\log g_{0_2} = \log \left( g_{0_1} g_{0_2} \right) = \log g_{1}  \\
    &= \frac{c_L}{3} \tanh^{-1}{\frac{l_L \tau_{3}}{2}}
    \end{split}
\end{equation}
where $\log g_{0_1}$ and $\log g_{0_2}$ are contributions from a single brane defect with tension $\tau_1$ and $\tau_2$ respectively~\cite{n,g}. 

Thus, the increment in entropy has a pleasingly simple form and agrees with field theory expectations: the $\log g$ values simply add.\footnote{In general, $\log\ g$ is additive under fusion of the branes, evidenced in how (\ref{oldLogG}) extends to (\ref{313}). However, it takes this simple form only when the AdS radii are the same.} As we introduce more freedom into the interface by looking at the fusion of multiple defects, the value of $g$ increases unbounded, indicating an infinite class if you want to interpret $g$ as classifying your defects. 

\smallskip
This result has to be compared and contrasted with our analogous results for the transmission coefficient \eqref{215}. There, we also found that the brane contributions add, but there it was the {\it tensions} that added. Due to the non-linear relation in terms of the $\tanh^{-1}$ function between $\log g$ and tension, this makes it very clear that merging two single brane defects gives us a genuinely new defect. We could, for example, try to fix the effective tension of a defect arising from the merger of two single brane interfaces with tensions $\tau_1$ and $\tau_2$ by requiring that the resulting $\log g$ should be that of a single brane with tension $\tau_3$ as shown in the last line of \eqref{314}. Re-expressing the $\tanh^{-1}$ in \eqref{314} as a $\log$ we find
\begin{equation}
    \tau_{3} = \frac{l_L(\tau_{1}+\tau_{2})}{1+\frac{l_L^2\tau_{1}\tau_{2}}{4}}
\end{equation}
If $\tau_3$ were just the sum $\tau_1 + \tau_2$, this would be consistent with our result for the transmission coefficient \eqref{215}. The extra non-linear term ruins the equivalence. 

\smallskip
Note that this even applies in the special case $l_{R}=l_{L}=l_{C}$ where we do not make use of the freedom to connect our original CFT to any ``auxilliary" CFT with a different central charge. We only look at interfaces from a given CFT to itself. Once again, looking at $\log g$ alone it appears that the system can be effectively described as a 2-region model with some tension $\tau_{3}$. But the transmission coefficient for a 2-region model with tension $\tau_{3}$ will not match the transmission coefficient expression obtained for the 3-region model.

\section{Conclusion and extensions}
In this work we have clearly demonstrated that in holographic bottom-up models for ICFTs based on RS branes, the fusion of two single brane interfaces does not yield back another single brane interface but instead should be thought of as a novel object in the theory. This somewhat alleviates the tension between the properties of the holographic toy interfaces and generic ICFTs: the 1-to-1 link between transmission coefficient and boundary entropy that was found in the single brane case is broken and the two can be independently dialed, at least over a certain range.

We only briefly touched upon the case of multiple interfaces. Clearly it would be of interest to consider the merger of not just two but of multiple branes. Presumably the list of all multi-brane interfaces will give a complete set of allowed interfaces.

Beyond that, it would be of much interest to compare and contrast our findings to those in a genuine top-down holographic model. As of now, the transmission coefficient has only been determined for the bottom-up scenario. $g$ has been calculated more widely. This will help determine whether the lack of fusion is indeed a generic feature inherent in the large $N$ limit underlying holographic constructions as we speculated in the introduction, or whether it is just a peculiarity of the particular bottom-up realization in terms of RS branes.

\acknowledgments
We appreciate helpful conversations with Costas Bachas, Ilka Brunner, Dongsheng Ge, and Marcos Riojas. This work was supported, in part, by the U.S.~Department of Energy under Grant DE-SC0022021 and by a grant from the Simons Foundation (Grant 651678, AK).

\appendix
\section{Appendix: Linear order differential equations and their solutions}
\label{appa}

This appendix contains the details of the linearized differential equations and their solutions and applied constraints for order 1 of $\epsilon$. 

\smallskip
Matching the metric gives the first 3 equations and the 4th one is from the non-diagonal element of the extrinsic curvature tensor's junction condition \eqref{eq:1}. The exact form of the 4th equation is arrived at after substituting one of the three equations to simplify the expression. 
\begin{equation}
    \begin{split}
        \Delta_{1}(z_{1}) + iwz_{1}\lambda_{1}(z_{1}) &= z_{1}^3 \left(  \frac{\cos{\theta_{L}}}{2} (I + R_{1}) - \frac{\cos{\theta_{CL}}}{2} (T_{1l}+R_{2l}) \right) \\
        \Delta_{1}(z_{1}) + z_{1}\partial_{z_{1}}\zeta_{1}(z_{1}) &= z_{1}^3 \left(  \frac{ \sin^2{\theta_{CL}} \cos{\theta_{CL}}}{2 }(T_{1l}+R_{2l})  - \frac{\sin^2{\theta_{L}} \cos{\theta_{L}}}{2}(I + R_{1})  \right) \\
        iwz_{1}\zeta_{1}(z_{1}) - z_{1}\partial_{z_{1}}\lambda_{1}(z_{1}) &= z_{1}^3 \left(\cos{\theta_{L}} \sin{\theta_{L}} (I - R_{1}) - \cos{\theta_{CL}} \sin{\theta_{CL}}(T_{1l} -R_{2l}) \right) \\
        z_{1}\partial_{z_{1}}D_{1}(z_{1}) &= z_{1}^3 \bigg( \frac{I - R_{1} - T_{1l} + R_{2l}}{iwz_{1}} + \frac{\cos^2{\theta_{CL}} \sin{\theta_{CL}}}{2}(T_{1l}+R_{2l}) \\
        &- \frac{\cos^2{\theta_{L}} \sin{\theta_{L}}}{2}(I + R_{1}) \bigg)
    \end{split}
\end{equation}
where $I$ and $R_{1}$ are the exponentials printed on the left brane and $T_{1l}$ and $R_{2l}$ the exponentials printed on the centre-left brane, given by:  
\begin{equation}
    \begin{split}
    I &= e^{-iw\sin{\theta_{L}}z_{1}}, \qquad \qquad R_{1} = \mathcal{R}_{L1} e^{iw\sin{\theta_{L}}z_{1}}, \\
    T_{1l} &= \mathcal{T}_{L1} e^{-iw\sin{\theta_{CL}}z_{1}}, \qquad R_{2l} = \mathcal{R}_{L2} e^{iw\sin{\theta_{CL}}z_{1}}. 
    \end{split}
\end{equation}
\begin{equation}
    \begin{split}
        \Delta_{2}(z_{2}) + iwz_{2}\lambda_{2}(z_{2}) &=  z_{2}^3 \left(  \frac{\cos{\theta_{CR}}}{2}(T_{1r} + R_{2r}) - \frac{\cos{\theta_{R}}}{2} T_{2} \right) \\
        \Delta_{2}(z_{2}) + z_{2}\partial_{z_{2}}\zeta_{2}(z_{2}) &= z_{2}^3 \left(  \frac{\sin^2{\theta_{R}} \cos{\theta_{R}}}{2}T_{2} - \frac{\sin^2{\theta_{CR}} \cos{\theta_{CR}}}{2} (T_{1r} + R_{2r}) \right) \\
        iwz_{2}\zeta_{2}(z_{2}) - z_{2}\partial_{z_{2}}\lambda_{2}(z_{2}) &= z_{2}^3 \left( \cos{\theta_{CR}} \sin{\theta_{CR}}(T_{1r} - R_{2r}) - \cos{\theta_{R}} \sin{\theta_{R}}T_{2} \right) \\
        z_{2}\partial_{z_{2}}D_{2}(z_{2}) &= z_{2}^3 \bigg( \frac{T_{1r} - R_{2r} - T_{2}}{iwz_{2}} - \frac{\cos{\theta_{CR}}^2 \sin{\theta_{CR}}}{2} (T_{1r}+R_{2r} ) \\
        &+ \frac{\cos^2{\theta_{R}} \sin{\theta_{R}}}{2}T_{2} \bigg)
    \end{split}
\end{equation}
where $T_{1r}$ and $R_{2r}$ are the exponentials printed on the center-right brane and $T_{2}$ the exponential printed on the right brane, given by:   
\begin{equation}
    \begin{split}
    T_{1r} &= \mathcal{T}_{L1} e^{-iw\sin{\theta_{CR}}z_{2}}, \qquad R_{2r} = \mathcal{R}_{L2} e^{iw\sin{\theta_{CR}}z_{2}}, \\
    T_{2} &= \mathcal{T}_{L1} e^{-iw\sin{\theta_{R}}z_{2}}. 
    \end{split}
\end{equation}

The $z_1 = 0$ and $z_2 = 0$ limits of these functions correspond to sources in the dual ICFT - displacing the interface or reparametrizing it. Since we are considering a non-fluctuating interface, we want to set the functions to 0 at the boundary (so for example $D_1(0) = 0 \rightarrow \delta_{L}(0) = \delta_{CL}(0)$ so there is no relative change in the interface). If you only look at the homogeneous equations where there are no gravitational waves scattered at the boundary, the solution looks like:
\begin{equation}\label{vacsol}
\begin{split} 
    \lambda_{1}(z_{1}) &= 
    \frac{i}{w} \left( a_{1+} e^{iwz_{1}} + a_{1-} e^{-iwz_{1}} \right) \\
    \zeta_{1}(z_{1}) &= \frac{i}{w} \left( a_{1+} e^{iwz_{1}} - a_{1-} e^{-iwz_{1}} \right) \\
    \Delta_{1}(z_{1}) &= z_{1} \left( a_{1+}e^{iwz_{1}} + a_{1-}e^{-iwz_{1}} \right) \\
    D_{1}(z_{1}) &= c_{1}
\end{split} 
\qquad
\begin{split}
    \lambda_{2}(z_{2}) &= 
    \frac{i}{w} \left( a_{2+} e^{iwz_{2}} + a_{2-} e^{-iwz_{2}} \right)\\
    \zeta_{2}(z_{2}) &= \frac{i}{w} \left( a_{2+} e^{iwz_{2}} - a_{2-} e^{-iwz_{2}} \right) \\
    \Delta_{2}(z_{2}) &= z_{2} \left( a_{2+}e^{iwz_{2}} + a_{2-}e^{-iwz_{2}} \right) \\
    D_{2}(z_{2}) &= c_{2}
\end{split}
\end{equation} 

Setting \eqref{vacsol} to 0 at the boundary, we get a constraint on the constants. Furthermore, it shows there are no solutions in the vacuum background supported purely by the interface: 
\begin{equation}
    \begin{split}
    c_1 &= 0 \qquad c_2 = 0  \\
    a_{1+} = 0 \qquad a_{1-} &= 0  \qquad
    a_{2+} = 0 \qquad a_{2-} = 0 
    \end{split}
\end{equation}
Using \eqref{vacsol}, the full solution to the full differential equations is: 
\begin{equation}
\begin{split}
    \lambda_{1}(z_{1}) &= 
    \frac{i}{w} \left( a_{1+} e^{iwz_{1}} + a_{1-} e^{-iwz_{1}} \right) + \frac{i}{w^{3}\cos{\theta_{L}}} \left( 1 - \frac{\cos^{2}{\theta_{L}} w^{2} z_{1}^{2}}{2}  \right) \left( I + R_{1} \right)  \\ 
    &- \frac{i}{w^{3}\cos{\theta_{CL}}}\left( 1 - \frac{\cos^{2}{\theta_{CL}} w^{2} z_{1}^{2}}{2} \right) (T_{1l}+ R_{2l}) \\
    \zeta_{1}(z_{1}) &= \frac{i}{w} \left( a_{1+} e^{iwz_{1}} - a_{1-} e^{-iwz_{1}} \right) 
   + \frac{i \tan{\theta_{CL}}}{w^3} \left( 1 + \frac{\cos^2{\theta_{CL}} w^{2} z_{1}^{2}}{2} \right) (T_{1l}-R_{2l})\\
    &- \frac{i \tan{\theta_{L}}}{w^3} \left( 1 + \frac{\cos^{2}{\theta_{L}} w^{2} z_{1}^{2}}{2} \right) (I - R_{1}) - \frac{z_{1}\cos{\theta_{L}}}{w^2}(I + R_{1})  + \frac{z_{1}\cos{\theta_{CL}}}{w^2}(T_{1l}+ R_{2l}) \\
    \Delta_{1}(z_{1}) &= z_{1} \left( a_{1+}e^{iwz_{1}} + a_{1-}e^{-iwz_{1}} + \frac{1}{w^{2} \cos{\theta_{L}}} (I + R_{1}) - \frac{1}{w^{2} \cos{\theta_{CL}}} (T_{1l}+ R_{2l})  \right) \\
    D_{1}(z_{1}) &= c_{1} - \frac{i}{w^{3}} \left( 1 + \frac{w^{2} z_{1}^{2} \cos^2{\theta_{L}}}{2} \right) (I - R_{1}) + \frac{i}{w^{3}} \left( 1 + \frac{w^{2} z_{1}^{2} \cos^2{\theta_{CL}}}{2} \right) (T_{1l} - R_{2l}) \\
    &+ \frac{z_{1}}{w^{2}} \left( \sin{\theta_{L}}(I + R_{1}) - \sin{\theta_{CL}}(T_{1l}+ R_{2l}) \right)
\end{split} 
\end{equation} 

\begin{equation}
\begin{split}
    \lambda_{2}(z_{2}) &= 
    \frac{i}{w} \left( a_{2+} e^{iwz_{2}} + a_{2-} e^{-iwz_{2}} \right) + \frac{i}{w^{3}\cos{\theta_{CR}}} \left( 1 - \frac{\cos^{2}{\theta_{CR}} w^{2} z_{2}^{2}}{2}  \right) \left(T_{1r} + R_{2r} \right)  \\ 
    &- \frac{i}{w^{3}\cos{\theta_{R}}}\left( 1 - \frac{\cos^{2}{\theta_{R}} w^{2} z_{2}^{2}}{2} \right) T_{2}\\
    \zeta_{2}(z_{2}) &= \frac{i}{w} \left( a_{2+} e^{iwz_{2}} - a_{2-} e^{-iwz_{2}} \right) 
   - \frac{i \tan{\theta_{CR}}}{w^3} \left( 1 + \frac{\cos^{2}{\theta_{CR}} w^{2} z_{2}^{2}}{2} \right) (T_{1r} - R_{2r})\\
    &+ \frac{i \tan{\theta_{R}}}{w^3} \left( 1 + \frac{\cos^2{\theta_{R}} w^{2} z_{2}^{2}}{2} \right)T_{2}  - \frac{z_{2}\cos{\theta_{CR}}}{w^2}(T_{1r} + R_{2r})  + \frac{z_{2}\cos{\theta_{R}}}{w^2}T_{2} \\
    \Delta_{2}(z_{2}) &= z_{2} \left( a_{2+}e^{iwz_{2}} + a_{2-}e^{-iwz_{2}} + \frac{1}{w^{2} \cos{\theta_{CR}}} (T_{1r} + R_{2r}) - \frac{1}{w^{2} \cos{\theta_{R}}} T_{2} \right) \\
    D_{2}(z_{2}) &= c_{2} - \frac{i}{w^{3}} \left( 1 + \frac{w^{2} z_{2}^{2} \cos^2{\theta_{CR}}}{2} \right)(T_{1r} - R_{2r}) + \frac{i}{w^{3}} \left( 1 + \frac{w^{2} z_{2}^{2} \cos^2{\theta_{R}}}{2} \right)T_{2}  \\
    &+ \frac{z_{2}}{w^{2}} \left( \sin{\theta_{CR}}(T_{1r} + R_{2r}) - \sin{\theta_{R}}T_{2} \right)
\end{split} 
\end{equation} 

For our non-fluctuating interface, we can again impose the sources go to 0 at the boundary. 

$\lambda_1(0)=0$ and $\zeta_1(0)=0$ give:
\begin{equation}
    \begin{split}
    a_{1+} = \frac{\sec{\theta_{CL}}(\mathcal{R}_{L2} + \mathcal{T}_{L1}) + \tan{\theta_{CL}}(\mathcal{R}_{L2} - \mathcal{T}_{L1}) - \sec{\theta_{L}}(1 + \mathcal{R}_{L1}) + \tan{\theta_{L}}(1 - \mathcal{R}_{L1})}{2w^2} \\
    a_{1-} = \frac{\sec{\theta_{CL}}(\mathcal{R}_{L2} + \mathcal{T}_{L1}) - \tan{\theta_{CL}}(\mathcal{R}_{L2} - \mathcal{T}_{L1}) - \sec{\theta_{L}}(1 + \mathcal{R}_{L1}) - \tan{\theta_{L}}(1 - \mathcal{R}_{L1})}{2w^2} 
    \end{split}
\end{equation}

$\lambda_2(0)=0$ and $\zeta_2(0)=0$ give:
\begin{equation}
    \begin{split}
    a_{2+} = \frac{-\sec{\theta_{CR}}(\mathcal{R}_{L2} + \mathcal{T}_{L1}) - \tan{\theta_{CR}}(\mathcal{R}_{L2} - \mathcal{T}_{L1}) + (\sec{\theta_{R}}-\tan{\theta_{R}}) \mathcal{T}_{L2})}{2w^2} \\
    a_{2-} = \frac{-\sec{\theta_{CR}}(\mathcal{R}_{L2} + \mathcal{T}_{L1}) + \tan{\theta_{CR}}(\mathcal{R}_{L2} - \mathcal{T}_{L1}) + (\sec{\theta_{R}}+\tan{\theta_{R}}) \mathcal{T}_{L2})}{2w^2} 
    \end{split}
\end{equation}

$a_{1+}$ and $a_{2+}$ correspond to the modes coming out of the horizon. These are unphysical and should be set to 0. Then, imposing $D_1(0)=0$, $D_2(0)=0$, $a_{1+}=0$ and $a_{2+}=0$ gives:
\begin{equation}
    \begin{split}
    1 &= \mathcal{R}_{L1} + \mathcal{T}_{L2} \\
    \mathcal{T}_{L2} &= \frac{2 \sec{\theta_{CR}} \sec{\theta_{L}}}{\sec{\theta_{CR}} (\sec{\theta_L} - \tan{\theta_{CL}}+ \tan{\theta_L}) + \sec{\theta_{CL}} (\sec{\theta_R} + \tan{\theta_{CR}} - \tan{\theta_R})} \\
    \mathcal{R}_{L2} &= \frac{\sec{\theta_L} \cos{\theta_{CL}}( \sec{\theta_{R}} -\sec{\theta_{CR}} + \tan{\theta_{CR}}- \tan{\theta_{R}})}{\sec{\theta_{R}} + \tan{\theta_{CR}}- \tan{\theta_{R}} - \sec{\theta_{CR}} \sin{\theta_{CL}} + \sec{\theta_{CR}} \cos{\theta_{CL}}(\sec{\theta_L} + \tan{\theta_L})} \\
    \mathcal{T}_{L1} &= \frac{\sec{\theta_L} \cos{\theta_{CL}}( \sec{\theta_{R}} + \sec{\theta_{CR}} + \tan{\theta_{CR}}- \tan{\theta_{R}})}{\sec{\theta_{R}} + \tan{\theta_{CR}}- \tan{\theta_{R}} - \sec{\theta_{CR}} \sin{\theta_{CL}} + \sec{\theta_{CR}} \cos{\theta_{CL}}(\sec{\theta_L} + \tan{\theta_L})}
    \end{split}
\end{equation}

\section{Appendix: Expressions for the minimal area lengths and log g}
\label{appb}

We obtain expressions for the minimal area lengths and correspondingly $\log g$ by imposing \eqref{34} for the case where the $M_{C}$ wedge is finite and found extending on both sides interface line. Based on the signs of the lengths of $R_L$ and $R_R$, we have constraints on $l_{L,R,C}$ and the tension bounds. 

\smallskip
For positive $R_L$, negative $R_R$ (all possible $l_{L,R,C}$):
\begin{equation}
    \begin{split}
        R_{L} &= \frac{1}{2} l_L \log \left(\frac{l_L^2-\left(\tau _1 l_C l_L+l_C\right){}^2}{l_C^2 \left(\tau _1 l_L-1\right){}^2-l_L^2}\right) \\
        R_{CL} &= -l_C \log \left(\frac{\left(\tau _1 l_C l_L+l_C+l_L\right) \left(l_C \left(\tau _1 l_L-1\right)+l_L\right)}{\sqrt{l_C^4 \left(-\left(\tau _1^2 l_L^2-1\right){}^2\right)+2 l_C^2 \left(\tau _1^2 l_L^4+l_L^2\right)-l_L^4}}\right) \\
        R_{CR} &= \frac{1}{2} l_C \log \left(\frac{\left(\tau _2 l_C l_R+l_C+l_R\right) \left(l_C \left(\tau _2 l_R-1\right)+l_R\right)}{\left(\tau _2 l_C \left(-l_R\right)+l_C+l_R\right) \left(\tau _2 l_C l_R+l_C-l_R\right)}\right)\\
        R_{R} &= -l_R \log \left(\frac{\left(\tau _2 l_C l_R+l_C\right){}^2-l_R^2}{\sqrt{l_C^4 \left(-\left(\tau _2^2 l_R^2-1\right){}^2\right)+2 l_C^2 \left(\tau _2^2 l_R^4+l_R^2\right)-l_R^4}}\right)
    \end{split}
\end{equation}
where 
\begin{equation}
    \sqrt{\left| \frac{1}{l_{L}^2} - \frac{1}{l_{C}^2}\right| } < \tau_{1} <  \frac{1}{l_{L}} + \frac{1}{l_{C}},\qquad  \qquad  \sqrt{\left| \frac{1}{l_{R}^2} - \frac{1}{l_{C}^2}\right| } < \tau_{2} < \frac{1}{l_{C}} + \frac{1}{l_{R}}
\end{equation}

\begin{equation} \label{logGbig}
    \begin{split}
    \log g &= \frac{1}{8 G} \left( 2 l_C \log \left(\frac{\left(\tau _1 l_C l_L-l_C+l_L\right) \left(\tau _1 l_C l_L+l_C+l_L\right)}{\sqrt{l_C^4 \left(-\left(\tau _1^2 l_L^2-1\right){}^2\right)+2 l_C^2 \left(\tau _1^2 l_L^4+l_L^2\right)-l_L^4}}\right) \right.\\
    &+ l_L \log \left(\frac{l_L^2-\left(\tau _1 l_C l_L+l_C\right){}^2}{l_C^2 \left(\tau _1 l_L-1\right){}^2-l_L^2}\right) + l_C \log \left(\frac{\left(\tau _2 l_C l_R-l_C+l_R\right) \left(\tau _2 l_C l_R+l_C+l_R\right)}{\left(\tau _2 l_C \left(-l_R\right)+l_C+l_R\right) \left(\tau _2 l_C l_R+l_C-l_R\right)}\right)\\
    &+ \left. 2 l_R \log \left(\frac{\left(\tau _2 l_C l_R+l_C\right){}^2-l_R^2}{\sqrt{l_C^4 \left(-\left(\tau _2^2 l_R^2-1\right){}^2\right)+2 l_C^2 \left(\tau _2^2 l_R^4+l_R^2\right)-l_R^4}}\right) \right)
    \end{split}
\end{equation}
For $l_L=l_R$, it turns out that this log $g$ expression is equivalent to the ones we express below (the different tension bounds and $l_{L,R,C}$ constraints still hold).

\smallskip
For negative $R_L$, positive $R_R$, ($l_{L,R,}>l_{C}$):
\begin{equation}
    \begin{split}
        R_{L} &= -\frac{1}{2} l_L \log \left(\frac{l_L^2-l_C^2 \left(\tau _1 l_L-1\right){}^2}{\left(\tau _1 l_C l_L+l_C\right){}^2-l_L^2}\right)\\ 
        R_{CL} &= -l_C \log \left(\frac{\left(\tau _1 l_C l_L+l_C+l_L\right) \left(l_C \left(\tau _1 l_L-1\right)+l_L\right)}{\sqrt{l_C^4 \left(-\left(\tau _1^2 l_L^2-1\right){}^2\right)+2 l_C^2 \left(\tau _1^2 l_L^4+l_L^2\right)-l_L^4}}\right)\\
        R_{CR} &= \frac{1}{2} l_C \log \left(\frac{\left(\tau _2 l_C l_R+l_C+l_R\right) \left(l_C \left(\tau _2 l_R-1\right)+l_R\right)}{\left(\tau _2 l_C \left(-l_R\right)+l_C+l_R\right) \left(\tau _2 l_C l_R+l_C-l_R\right)}\right)\\
        R_{R} &= l_R \log \left( \frac{ \left| l_R^2-l_C^2 \left(l_R^2 \tau _2^2+1\right) \right| +2 \tau _2 l_C^2 l_R }{\sqrt{-\left( l_C^4 \left(\tau _2^2 l_R^2-1\right){}^2-2 l_C^2 \left(\tau _2^2 l_R^4+l_R^2\right)+l_R^4 \right)}} 
        \right)
    \end{split}
\end{equation}

where 
\begin{equation}
    \frac{1}{l_{C}} - \frac{1}{l_{L}} < \tau_{1} < \sqrt{\frac{1}{l_{C}^2} - \frac{1}{l_{L}^2}},\qquad  \qquad  \frac{1}{l_{C}} - \frac{1}{l_{R}} < \tau_{2} < \sqrt{\frac{1}{l_{C}^2} - \frac{1}{l_{R}^2}}    
\end{equation}

\begin{equation}
\begin{split}
\log g &= \frac{1}{8 G} \left( -2 l_R \log \left( \frac{ \left| l_R^2-l_C^2 \left(l_R^2 \tau _2^2+1\right) \right| +2 \tau _2 l_C^2 l_R }{\sqrt{-\left( l_C^4 \left(\tau _2^2 l_R^2-1\right){}^2-2 l_C^2 \left(\tau _2^2 l_R^4+l_R^2\right)+l_R^4 \right)}} 
        \right) \right. \\
&+ l_C \log \left(\frac{\left(\tau _2 l_C l_R-l_C+l_R\right) \left(\tau _2 l_C l_R+l_C+l_R\right)}{\left(\tau _2 l_C \left(-l_R\right)+l_C+l_R\right) \left(\tau _2 l_C l_R+l_C-l_R\right)}\right) -l_L \log \left(\frac{l_L^2-l_C^2 \left(\tau _1 l_L-1\right){}^2}{\left(\tau _1 l_C l_L+l_C\right){}^2-l_L^2}\right) \\
&+ \left. 2 l_C \log \left(\frac{\left(\tau _1 l_C l_L-l_C+l_L\right) \left(\tau _1 l_C l_L+l_C+l_L\right)}{\sqrt{l_C^4 \left(-\left(\tau _1^2 l_L^2-1\right){}^2\right)+2 l_C^2 \left(\tau _1^2 l_L^4+l_L^2\right)-l_L^4}}\right) \right) 
\end{split}
\end{equation}

For negative $R_L$, negative $R_R$ ($l_{L}>l_{C}$), we can obtain similar expressions for $R_{L}$, $R_{CL}$, $R_{CR}$, $R_{R}$ and Log $g$ but, perhaps more interestingly, the tension bounds are given by:
\begin{equation}
    \frac{1}{l_{C}} - \frac{1}{l_{L}} < \tau_{1} < \sqrt{\frac{1}{l_{C}^2} - \frac{1}{l_{L}^2}},\qquad  \qquad  \sqrt{\left| \frac{1}{l_{R}^2} - \frac{1}{l_{C}^2}\right|} < \tau_{2} <  \frac{1}{l_{C}} + \frac{1}{l_{R}}   
\end{equation}

\smallskip
Likewise, for positive $R_L$, positive $R_R$ ($l_{R}>l_{C}$), we can obtain similar expressions for $R_{L}$, $R_{CL}$, $R_{CR}$, $R_{R}$ and Log $g$ but now the tension bounds are:
\begin{equation}
     \sqrt{\left| \frac{1}{l_{L}^2} - \frac{1}{l_{C}^2}\right|} < \tau_{1} <  \frac{1}{l_{C}} + \frac{1}{l_{L}} ,\qquad  \qquad \frac{1}{l_{C}} - \frac{1}{l_{R}} < \tau_{2} < \sqrt{\frac{1}{l_{C}^2} - \frac{1}{l_{R}^2}}
\end{equation}

\bibliographystyle{JHEP}
\bibliography{ref}

\end{document}